\documentclass[11pt]{article}
\usepackage{amsmath}
\usepackage{amsfonts}
\usepackage[dvips]{epsfig}

 \advance \textwidth by -2in
 \advance \textheight by -3in
\def\##1{{\bf #1}}
\def\=#1{\underline{\underline{#1}}}

\def\+#1{\underline{\bf #1}}
\def\*#1{\underline{\underline{\bf #1}}}

\def\eps{\epsilon}
\def\epso{\epsilon_0}
\def\muo{\mu_0}
\def\ko{k_0}

\def\.{\mbox{ \tiny{$^\bullet$} }}

\def\le{\left(}
\def\ri{\right)}
\def\les{\left[}
\def\ris{\right]}
\def\lec{\left\{}
\def\ric{\right\}}

\def\c#1{\cite{#1}}

\begin{document}

\noindent {\bf COUNTERPOSITION AND NEGATIVE REFRACTION DUE TO UNIFORM MOTION  }
\vskip 0.2cm

\noindent  {\bf Tom G. Mackay$^1$ and Akhlesh Lakhtakia$^2$} \vskip
0.2cm

\noindent {\sf $^1$ School of Mathematics\\
\noindent University of Edinburgh\\
\noindent Edinburgh EH9 3JZ, United Kingdom} \vskip 0.4cm

\noindent {\sf $^2$ CATMAS~---~Computational \& Theoretical Materials Sciences Group \\
\noindent Department of Engineering Science \& Mechanics\\
\noindent 212 Earth \& Engineering Sciences Building\\
\noindent Pennsylvania State University, University Park, PA
16802--6812} \vskip 0.4cm

\

\noindent {\bf ABSTRACT:} Refraction of obliquely incident plane
waves due to the interface of a vacuous half--space and  a
half--space occupied by a simply moving, nondissipative, isotropic
dielectric--magnetic medium is considered, when the medium's
velocity lies parallel to the interface and in the plane of
incidence.  Counterposition of the refracted wavevector and
time--averaged Poynting vector occurs when the medium's velocity is
sufficiently large in magnitude and makes an obtuse angle with the
incident wavevector.  The counterposition induced by relative motion
occurs whether the refraction is negative or positive when the
medium is at rest.

 \vskip 0.4cm

\vspace{10mm}

\noindent{\bf 1. INTRODUCTION}

When a plane wave is incident upon the planar interface of two
homogeneous mediums, the refracted wavevector and time--averaged
Poynting vector can emerge on opposite sides of the  normal to
the interface. This phenomenon is called \emph{counterposition}.
Conditions for the occurrence of counterposition in uniaxial dielectric--magnetic
mediums have been established  \c{Optik}.
Under the title of amphoteric refraction, counterposition has been
confused with negative refraction \c{Zhang}.
Refraction~---~whether negative or positive~---~concerns the orientation of the refracted
wavevector relative to the normal to the interface, as per a law often attributed
to Willebrord van Snel van Royen
\c{Chen}. The orientation of the refracted time--averaged Poynting
vector is irrelevant to whether or not the refraction is negative or
positive.

In this Letter, we establish the prospects for counterposition due
to the interface of a vacuous half--space and  a half--space
occupied by a simply moving, nondissipative, isotropic
dielectric--magnetic medium, when the medium's velocity lies
parallel to the interface and in the plane of incidence. In
particular, we show that counterposition and negative refraction are
indeed distinct. For background details of planewave propagation in
moving mediums, the reader is referred to standard works
\c{Chen,Pappas,ChUn}.

\noindent{\bf 2. ANALYSIS}

Suppose that a plane wave is launched with wavevector
$\mbox{\boldmath$k_i$} = k_i \mbox{\boldmath$\hat{k}_i$}$ from
vacuum towards a half--space occupied by an isotropic,
nondissipative, dielectric--magnetic medium. Let this medium move at
constant velocity $\mbox{\boldmath$v$} = v \mbox{\boldmath$\hat{v}$}
$, parallel to the interface and in the plane of incidence. With
respect to an inertial frame of reference that moves with the
same velocity $\mbox{\boldmath$v$}$ with respect to
the laboratory frame of reference wherein $\mbox{\boldmath$k_i$}$
is specified, the refracting medium is
characterized by relative permittivity $\eps_r  $ and relative
permeability $\mu_r $.
  The condition $\eps_r \mu_r \ge 1$ is
assumed  in order to exclude the possibility of
evanescent plane waves.

The angle $\phi_t$ between the refracted
wavevector $\mbox{\boldmath$k_t$} = k_t
\mbox{\boldmath$\hat{k}_t$}$, as observed from the laboratory
frame of reference, and the unit vector
$\mbox{\boldmath$\hat{q}$}$ normal to the interface is related to the angle of incidence
\begin{equation}
\phi_i = \cos^{-1} \le \mbox{\boldmath$\hat{k}$}_i
\.\mbox{\boldmath$\hat{q}$}
   \ri
\end{equation}
 by \c{Chen}
\begin{equation}
  \phi_t = \sin^{-1}  \le \frac{\ko \sin
\phi_i}{k_t} \ri \,,
\end{equation}
where
\begin{equation}
  k_t =
\left\{
\begin{array}{lcr}
  \ko \lec 1 + \xi \les 1 -
\beta \le \mbox{\boldmath$\hat{k}_i$} \. \mbox{\boldmath$\hat{v}$}
\ri \ris^2 \ric^{1/2} & \mbox{for} & \eps_r, \mu_r > 0
\\ && \\
  -\ko \lec 1 + \xi \les 1 -
\beta \le \mbox{\boldmath$\hat{k}_i$} \. \mbox{\boldmath$\hat{v}$}
\ri \ris^2 \ric^{1/2} & \mbox{for} & \eps_r, \mu_r < 0
\end{array}\right.
\end{equation}
is the wavenumber of the refracted wave,
  $\ko$ is the wavenumber in vacuum,
\begin{equation}
   \xi =
\frac{\eps_r \mu_r - 1}{1 - \beta^2}\,,
\end{equation}
 and $\beta = v \sqrt{\epso
\muo}$ with $\epso$ and $\muo$ being the permittivity and
permeability of vacuum.

\begin{figure}[!ht]
\centering \psfull \epsfig{file=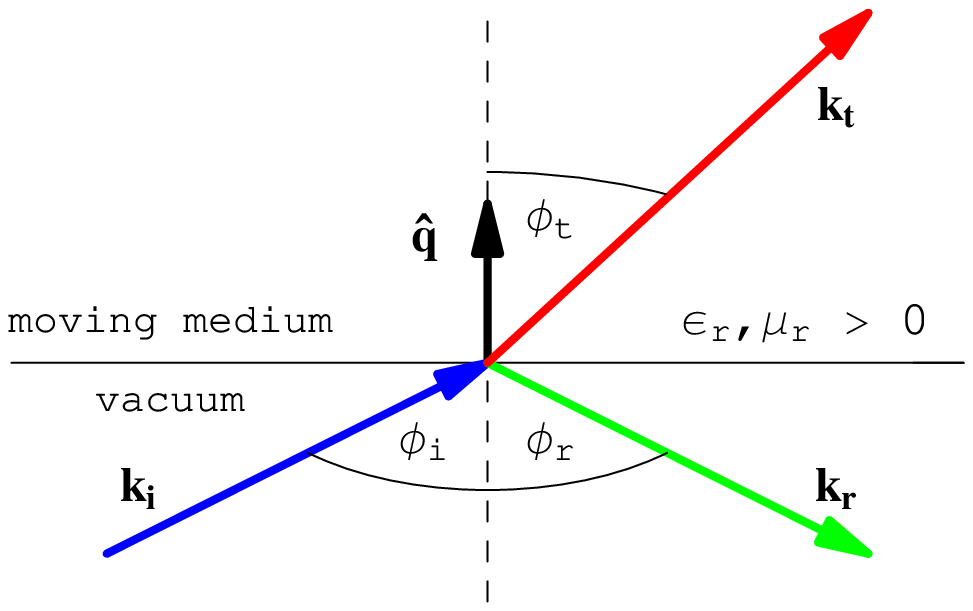,width=4.0in}
  \caption{\label{fig1} A plane wave with wavevector $\mbox{\boldmath$k_i$}$
  is incident from vacuum onto a half--space occupied by a simply moving medium at an angle  $\phi_i$ with
respect to the unit vector $\mbox{\boldmath$\hat{q}$}$ normal to the
planar interface. The moving medium is characterized by relative
permittivity $\eps_r > 0$ and relative permeability $\mu_r > 0$ in a
comoving frame of reference. As observed in the non--comoving
(laboratory) frame
of reference wherein the incident plane wave is specified, the
refracted wavevector $\mbox{\boldmath$k_t$}$ makes an angle $\phi_t$
with $\mbox{\boldmath$\hat{q}$}$.
  }
\end{figure}

\begin{figure}[!ht]
\centering \psfull \epsfig{file=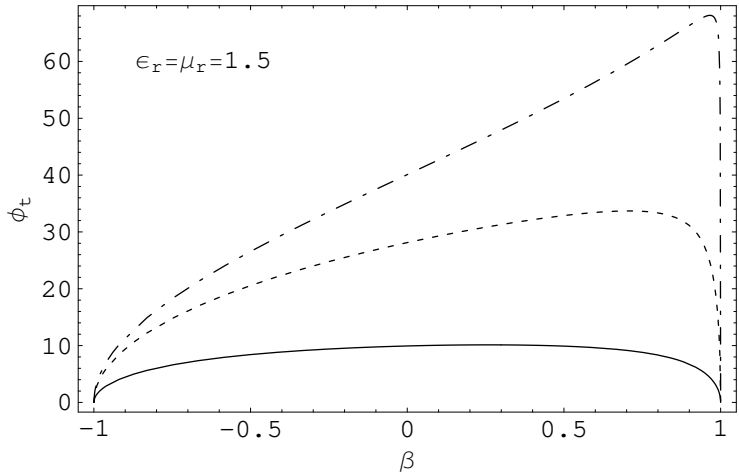,width=4.0in}
  \caption{\label{fig2} The angle of refraction $\phi_t$ (in degree)
   plotted as a function of $\beta \in (-1,1)$, when the angle of
incidence $\phi_i = 15^\circ$ (solid curves), $45^\circ$ (dashed
curves) and $75^\circ$ (broken dashed curves); $\eps_r = \mu_r =1.5$.
  }
\end{figure}

Let us consider case where   $\eps_r $ and $\mu_r $ are
positive--valued. Then,  $0 < \phi_t < \pi/2$ for all $\phi_i \in
(0, \pi/2)$. That is, the refraction is always positive
$\forall\,\beta \in (-1,1)$, as is illustrated schematically in
Fig.~\ref{fig1}.  Plots of $\phi_t$ against $\beta \in (-1,1)$ for
three values of $\phi_i$ are provided in Fig.~\ref{fig2}.

\begin{figure}[!ht]
\centering \psfull \epsfig{file=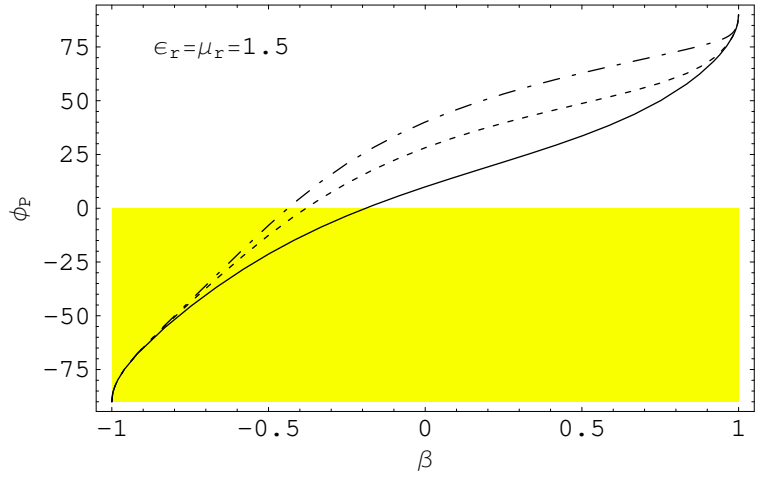,width=4.0in}
  \caption{\label{fig3} As Figure~\ref{fig2}, but for the
  angle $\phi_P$ (in degree) between the refracted time--averaged
  Poynting vector $\mbox{\boldmath$\hat{P}$}_t$ and
  the unit vector $\mbox{\boldmath$\hat{q}$}$.
  The counterposition regime  $\phi_P < 0^\circ$ is shaded.
  }
\end{figure}

The time--averaged
Poynting vector of the refracted plane wave is given by \c{Chen}
\begin{equation}
\mbox{\boldmath$P$}_t = P_t \, \mbox{\boldmath$\hat{P}$}_t = \le
\frac{1}{\mu_r} \,| C_1 |^2 +  \eps_r  | C_2 |^2 \ri \le
\mbox{\boldmath$k_t$} \times \mbox{\boldmath$\hat{v}$} \ri^2 \les
\mbox{\boldmath$k_t$} + \xi \beta  \le \ko - \beta
\mbox{\boldmath$k_t$} \. \mbox{\boldmath$\hat{v}$} \ri
\mbox{\boldmath$\hat{v}$} \ris,
\end{equation}
where $C_{1}$ and $C_2$ are constants. The angle
\begin{equation}
 \phi_P = \tan^{-1} \le \frac{\mbox{\boldmath$\hat{P}$}_t \.\mbox{\boldmath$v$}}{
| v | \, \mbox{\boldmath$\hat{P}$}_t \.\mbox{\boldmath$\hat{q}$}}
   \ri
\end{equation}
 between $\mbox{\boldmath$\hat{q}$}$ and
$\mbox{\boldmath$\hat{P}$}_t$ is plotted in Fig.~\ref{fig3} against
$\beta \in (-1,1)$  for three values of $\phi_i$. The orientation of
the refracted time--averaged Poynting vector rotates towards the   direction
of motion as $\beta$ increases from $-1$. Clearly, counterposition arises
$\forall\,\beta < \tilde\beta$ where $\tilde\beta$ is some negative number that
depends, among other quantities, on $\phi_i$.

\begin{figure}[!ht]
\centering \psfull \epsfig{file=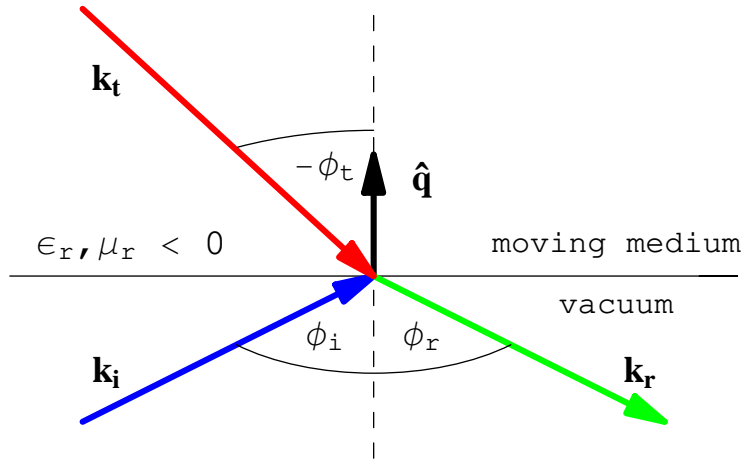,width=4.0in}
  \caption{\label{fig4} As Figure~\ref{fig1} but for $\eps_r < 0$ and
$\mu_r < 0$.
  }
\end{figure}

\begin{figure}[!ht]
\centering \psfull \epsfig{file=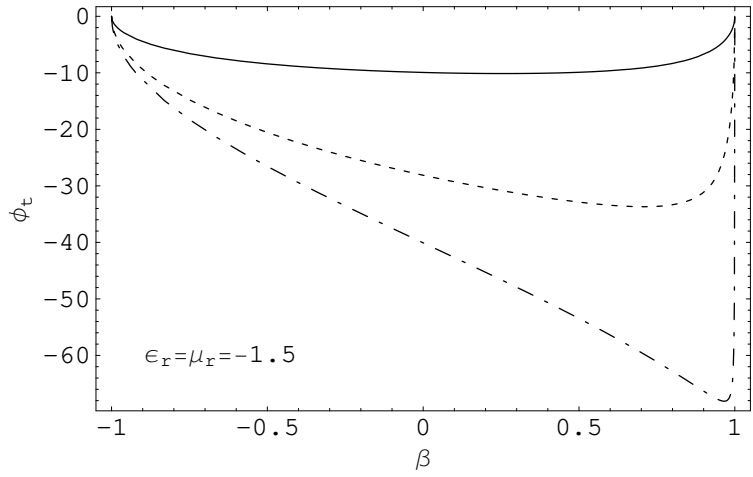,width=4.0in}
  \caption{\label{fig5} As Figure~\ref{fig2} but for  $\eps_r = \mu_r =-1.5$.
  }
\end{figure}

Now let us turn to the  scenario wherein $\eps_r < 0 $ and $\mu_r <
0$,  as schematically represented in Fig.~\ref{fig4}. As a
consequence of both $\eps_r$ and $\mu_r $ being negative--valued, we
have $- \pi/2 < \phi_t < 0 $ for all $\phi_i \in (0, \pi/2)$. That
is, the refraction is always negative $\forall\,\beta \in (-1,1)$.
The plots in Fig.~\ref{fig5} of $\phi_t$ against $\beta \in (-1,1)$
for three values of $\phi_i$ illustrate that conclusion.

\begin{figure}[!ht]
\centering \psfull \epsfig{file=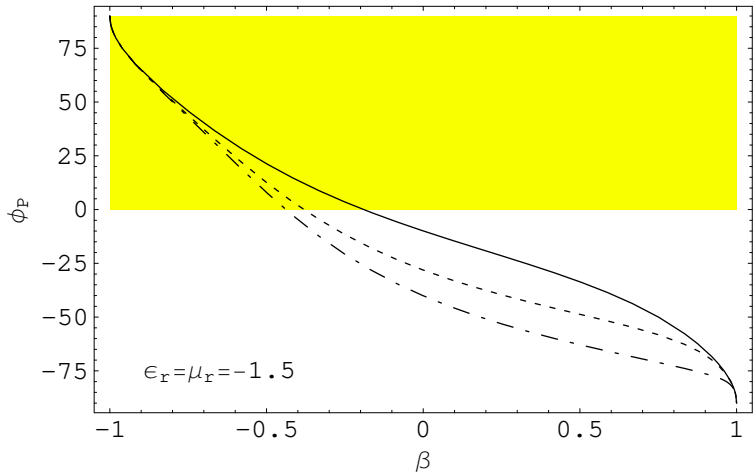,width=4.0in}
  \caption{\label{fig6} As Figure~\ref{fig3} but for  $\eps_r = \mu_r =-1.5$.  The counterposition regime  $\phi_P > 0^\circ$ is shaded.
  }
\end{figure}

The corresponding orientation angles for the refracted time--averaged Poyting
vector are graphed against $\beta \in (-1,1)$ in Fig.~\ref{fig6}. As
is the case for the positively refracting scenario, we see that
counterposition arises $\forall\,\beta < \tilde\beta$ where
$\tilde\beta$ is negative. However, in contrast to the positively
refracting scenario,  the refracted
time--averaged Poynting vector rotates against the  direction of
motion as $\beta$ increases.

\noindent{\bf 3. CONCLUDING REMARKS}

Thus, counterposition may be induced in an isotropic
dielectric--magnetic medium by relative motion at constant
velocity~---~whether the medium is positively or negatively
refracting when at rest. Thereby, the distinction between
counterposition and negative refraction is further emphasized
\c{Zhang,GK}. Also, in the scenario considered here, we note that
the phase velocity of the refracted plane wave is positive when
positive refraction occurs and negative when negative refraction
occurs. Hence,  negative phase velocity is not induced by relative
motion parallel to the medium interface \c{NPV_STR}. In the absence
of relative motion (i.e., $\beta = 0$) the refracted wavevector and
time--averaged Poynting vector are parallel for $\eps_r, \mu_r > 0$
and anti--parallel for $\eps_r, \mu_r < 0$, and accordingly
counterposition cannot occur, as is confirmed by Figs.~\ref{fig2},
\ref{fig3}, \ref{fig5} and \ref{fig6}.

\vspace{10mm}

\noindent{\bf Acknowledgement:}  TGM is supported by a \emph{Royal
Society of Edinburgh/Scottish Executive Support Research
Fellowship}.

\newpage

\vspace{5mm}

\vspace{5mm}

\vspace{5mm}

\vspace{5mm}

\end{document}